\documentclass{cjpsuplf}    
\usepackage[dvips]{graphicx}
\begin{document}
\title{Polarized Photon Structure:\ $g_1^\gamma$ and $g_2^\gamma$
\,\footnote{
Presented by T. Uematsu at the Advanced Study Institute on Symmetries
and Spin, Praha-SPIN 2001, Prague, July 15-28, 2001.
KUCP-203, YNU-HEPTh-02-103, to appear in the Proceedings.
}}
\authori{Hideshi Baba}
\addressi{Graduate School of Human and Environmental Studies,
Kyoto University \\ Kyoto 606-8501, Japan}
\authorii{Ken Sasaki}
\addressii{Dept. of Physics, Faculty of Engineering, Yokohama
National University \\ Yokohama 240-8501, Japan}
\authoriii{Tsuneo Uematsu}     
\addressiii{Dept. of Fundamental Sciences, FIHS, Kyoto University
\\ Kyoto 606-8501, Japan}
\headtitle{Polarized photon structure \ldots}
\headauthor{Hideshi Baba et al.}  
\specialhead{Hideshi Baba et al.: Polarized photon structure  \ldots}
\evidence{}
\daterec{}    
\suppl{A}  \year{2002}
\setcounter{page}{1}
\maketitle

\begin{abstract}

We investigate the polarized photon structure functions $g_1^\gamma$ and
$g_2^\gamma$ which can be studied in the future polarized version of
ep or e$^+$e$^-$ colliders. The NLO QCD calculations of $g_1^\gamma$ and
the possible twist-3 effects in $g_2^\gamma$ are discussed.

\end{abstract}

\section{Introduction}     

In the last ten years, the spin-dependent structure functions $g_1$ 
and $g_2$ of the nucleon have been intensively studied in the polarized
deep-inelastic scattering of polarized lepton on polarized nucleon 
targets. Based on the next-to-leading order (NLO) QCD analysis of the 
experimental data, the polarized parton distributions inside the 
nucleon have been extracted. The gluon polarization $\Delta G$ is now
one of the central issues in spin physics.

On the other hand, there has been growing interest in the 
polarized photon structure functions. Especially the first moment of
a photon structure function $g_1^\gamma$ has attracted much attention 
in the literature \cite{BASS,ET,NSV,FS,BBS} 
in connection with the axial anomaly which is also 
relevant in the nucleon spin structure function $g_1^{p(n)}$. 
The polarized structure function $g_1^\gamma$ could be experimentally
studied in the polarized version of ep collider HERA \cite{Barber,SVZ}, 
or more directly measured by the polarized e$^+$e$^-$ collision in
the future linear collider. And the next-to-leading order QCD analysis
of $g_1^\gamma$ has been performed in the literature \cite{SV,SU,SU2,GRS}.

Now there exists another structure function $g_2^\gamma$ for the virtual
photon target, where the twist-3 effect is also relevant in addition to
the usual twist-2 effect. In this talk, we first briefly summarize our
results for the $g_1^\gamma$ for the virtual photon target, and then 
investigate the twist-3 effects in $g_2^\gamma$.

\section{$g_1^\gamma(x,Q^2,P^2)$ and QCD Sum Rule}

We consider the polarized deep inelastic scattering on a polarized virtual
photon target and study the virtual photon structure functions 
for the kinematical region:
\bea
\Lambda^2 \ll P^2 \ll Q^2,
\eea
where $-Q^2$ ($-P^2$) is the mass squared of the probe (target) photon,
and $\Lambda$ is the QCD scale parameter.

The same framework used in the analysis of nucleon spin
structure functions can be applied to the present case.
Namely, we can either base our argument on the operator
product expansion (OPE) with the use of renormalization group
method (RG) or on the DGLAP parton evolution equations.

The $n$-th moment of $g_1^\gamma(x,Q^2,P^2)$ turns out to
be given by \cite{SU}:
\begin{eqnarray}
&&\int_0^1 dx x^{n-1}g_1^\gamma(x,Q^2,P^2)  \nonumber\\
&=&\frac{\alpha}{4\pi}\frac{1}{2\beta_0}
\left[\sum_{i=+,-,NS}
L_i^n
\frac{4\pi}{\alpha_s(Q^2)}
\left\{1-\left(\frac{\alpha_s(Q^2)}{\alpha_s(P^2)}\right)^{\lambda_i^n/2\beta_0
+1}\right\}\right.\nonumber\\
&&\hspace{2.0cm}+\left.\sum_{i=+,-,NS}{\cal A}_i^n\left\{1-\left(\frac{\alpha_s(Q^2)}{\alpha_s(P^2)}\right)^{\lambda_i^n/2\beta_0}\right\}\right.\nonumber\\
&&\hspace{2.0cm}+\left.\sum_{i=+,-,NS}{\cal B}_i^n\left\{1-\left(\frac{\alpha_s(Q^2)}{\alpha_s(P^2)}\right)^{\lambda_i^n/2\beta_0+1}\right\}\right.\nonumber\\
&&\vspace{2cm}\hspace{5.0cm}+\left.{\cal C}^n +{\cal O}(\alpha_s) \ \right],
\label{master}
\end{eqnarray}
where $L_i^n$, ${\cal A}_i^n$, ${\cal B}_i^n$ and ${\cal C}^n$
are computed from the NLO QCD perturbation theory.
$\lambda_i^n \ (i=+,-,NS)$ are eigenvalues of one-loop anomalous 
dimensions. Note that the same formula with the different coefficients holds
for the unpolarized structure function $F_2^\gamma(x,Q^2,P^2)$ \cite{UW}.
We also note that the real photon's $g_1^\gamma$ was studied to the LO
in \cite{S} and to the NLO in \cite{SV}.

One of the remarkable consequences from the above results is the 
non-vanishing first moment of $g_1^\gamma(x,Q^2,P^2)$, for which 
we have
\bea
\int_0^1 dx g_1^\gamma(x,Q^2,P^2)=
-\frac{3\alpha}{\pi}\sum_{i=1}^{n_f}{e_i}^4
+{\cal O}(\alpha_s), \label{lowest}
\eea
in contrast to the vanishing first moment for the real photon case ($P^2=0$):
\bea
\int_0^1 dx g_1^\gamma(x,Q^2)=0,
\eea
which holds to all orders of $\alpha_s(Q^2)$ in QCD \cite{BBS}.
We have also computed the ${\cal O}(\alpha_s)$ 
QCD corrections in (\ref{lowest})\cite{SU}, which coincides 
with the result obtained in \cite{NSV}.

\section{Spin-Dependent Parton Distributions}
Now the factorization theorem tells us that the physically observable
quantities like cross sections or structure functions can be
factored into the long-distance part (distribution function) and
short-distance part (coefficient function). Thus the polarized 
photon structure function can be schematically written as
\begin{equation}
g_1^\gamma=\Delta\vec{q}^\gamma \ \otimes\  \Delta\vec{C}^\gamma,
\label{factorization}
\end{equation}
where spin-dependent parton distributions $\Delta\vec{q}^\gamma$:
\begin{equation}
\Delta\vec{q}^\gamma(x,Q^2,P^2)=(\Delta q_S^\gamma,\Delta
 G^\gamma, \Delta q_{NS}^\gamma, \Delta \Gamma^\gamma)
\end{equation}
are polarized flavor-singlet quark, gluon, non-singlet quark 
and photon distribution functions in the virtual photon
(we put the symbol $\Delta$ for polarized quantities),
and  
\begin{equation}
{\Delta\vec{C}^\gamma}^T=(\Delta C_S^\gamma,\ \Delta C_G^\gamma,\ 
\Delta C_{NS}^\gamma, \ \Delta C_\gamma^\gamma)
\end{equation}
are the corresponding coefficient functions.   
In the leading order in QED coupling $\alpha=\frac{e^2}{4\pi}$, the
photon distribution function can be taken as 
$\Delta \Gamma^\gamma (x,Q^2,P^2)=\delta(1-x)$. Therefore
we have the following inhomogeneous DGLAP evolution equation for
$\Delta\mbox{\boldmath $q$}^\gamma=(\Delta q_S^\gamma,\Delta
 G^\gamma, \Delta q_{NS}^\gamma)$:
\begin{equation}
\frac{d\Delta\mbox{\boldmath $q$}^\gamma(x,Q^2,P^2)}{d\ln Q^2}=
\Delta\mbox{\boldmath $K$}(x,Q^2)+\int_x^1\frac{dy}{y}
\Delta\mbox{\boldmath $q$}^\gamma
(y,Q^2,P^2)\times \Delta P(\frac{x}{y},Q^2),\label{DGLAP}
\end{equation}
where $\Delta\mbox{\boldmath $K$}(x,Q^2)$ is the splitting function of 
the photon into quark and gluon, whereas $\Delta P({x}/{y},Q^2)$ is 
the 3$\times$3 splitting function matrix.
The solution to the DGLAP evolution equation can be given by
\begin{equation}
\Delta\vec{q}^\gamma(t)=\Delta\vec{q}^{\gamma(0)}(t)
+\Delta\vec{q}^{\gamma(1)}(t),\quad 
t \equiv \frac{2}{\beta_0}\ln \frac{\alpha_s(P^2)}{\alpha_s(Q^2)},
\end{equation}
where the first (second) term corresponds to LO (NLO) approximation.
The initial condition we impose is the following,
\begin{equation}
\Delta\vec{q}^{\gamma(0)}(0)=0, \quad 
\Delta\vec{q}^{\gamma(1)}(0)=\frac{\alpha}{4\pi}\vec{A}_n,
\end{equation}
where $\vec{A}_n$ is the constant which depends on the factorization
scheme to be used. 
Or equivalently in the language of OPE, this constant appears
as a finite matrix element of the operators, $\vec{O}_n$ renormalized at
$\mu^2=P^2$  
between the photon states:
\begin{equation}
\langle \gamma (p) \mid \vec{O}_n (\mu) \mid \gamma (p) 
\rangle|_{\mu^2=P^2} =\frac{\alpha}{4\pi}\vec{A}_n.
\end{equation}
This scheme dependence arises from the freedom of multiplying the arbitrary
finite renormalization constant $Z_a$ and its inverse $Z_a^{-1}$
in the $n$-th moment of (\ref{factorization}):
\begin{eqnarray}
g_1^\gamma(n,Q^2,P^2)=\Delta\vec{q}^\gamma\cdot\Delta\vec{C}^\gamma
=\Delta\vec{q}^\gamma Z_a \cdot 
Z_a^{-1}\Delta\vec{C}^\gamma
=\Delta\vec{q}^\gamma|_a \cdot \Delta\vec{C}^\gamma|_a,
\end{eqnarray}
where the resulting $\Delta\vec{q}^\gamma|_a$ and $\Delta\vec{C}^\gamma|_a$
are the distribution function and the coefficient function in the $a$-scheme.
For the parton distributions in various schemes, see ref.\cite{SU2}.

\section{$g_2^\gamma(x,Q^2,P^2)$ and Twist-3 Effects}

The antisymmetric part of the structure tensor, $W^A_{\mu\nu\rho\tau}(p,q)$
for the target (probe) photon with momentum $p$ ($q$), 
relevant for the polarized structure, can be written in terms of the structure
functions, $g_1^\gamma$ and $g_2^\gamma$, as \cite{SU}
\bea
W^A_{\mu\nu\rho\tau}=\frac{1}{(p\cdot q)^2}
[(I_{-})_{\mu\nu\rho\tau}\ g_1^\gamma
-(J_{-})_{\mu\nu\rho\tau}\ g_2^\gamma] ,\ 
\eea
where the two tensor structures are explicitly given by
\bea
(I_{-})_{\mu\nu\rho\tau}&\equiv&p\cdot q\,\epsilon_{\mu\nu\lambda\sigma}
{\epsilon_{\rho\tau}}^{\sigma\beta}q^\lambda p_\beta,\\
&& {}\nonumber\\
(J_{-})_{\mu\nu\rho\tau}&\equiv& \epsilon_{\mu\nu\lambda\sigma}
\epsilon_{\rho\tau\alpha\beta}q^\lambda p^\sigma q^\alpha
p^\beta-p\cdot q\,\epsilon_{\mu\nu\lambda\sigma}
{\epsilon_{\rho\tau}}^{\sigma\beta}q^\lambda p_\beta.
\eea
Here we note that we have the structure function $g_2^\gamma$
only for non-zero $P^2$, i.e. virtual photon. 
For the nucleon, the twist-3 contribution to $g_2$ is
negligibly small in the experimental data sor far obtained \cite{E143,E155}.

Now let us decompose $g_2$ into twist-2 and twist-3 contributions:
\bea
g_2=g_2^{\rm tw.2}+g_2^{\rm tw.3}.
\eea
Experimental data for nucleons show
\bea
g_2 \approx g_2^{\rm tw.2}=g_2^{\rm WW},
\eea
where $g_2^{\rm WW}$ is Wandzura-Wilczek relation \cite{WW}:
\bea
g_2^{\rm WW}(x,Q^2)=-g_1(x,Q^2)+\int_x^1
\frac{dy}{y}g_1(y,Q^2).
\eea
Now we ask what about the photon structure, especially virtual
photon case ?

First we consider the operator product expansion (OPE) relevant for the 
photon structure functions. The OPE can be decomposed into the twist-2
and twist-3 contributions as follows:
\bea
\int d^4 x e^{iq\cdot x}
J_{\mu}(x)J_{\nu}(0)^{[A]} \sim\sum_{n} R_n^{(2)}(0)E_n^{(2)}(Q^2)
+ \sum_{n} R_n^{(3)}(0)E_n^{(3)}(Q^2),
\eea
where $R_n^{(2)}$ and $R_n^{(3)}$ denote the twist-2 and twist-3
operators, respectively. For the nucleon the matrix element:
\bea
\langle N(p,s)|R_n^{(3)}|N(p,s)\rangle
\eea
is small. For the photon, as we will see below, the matrix
element is non-vanishing:
\bea
\langle\gamma(p,s)|R_n^{(3)}|\gamma(p,s)\rangle\neq 0,
\eea
which contributes to $g_2^\gamma$ for the virtual photon. Namely we have
\bea
&&\int d^4 x e^{iq\cdot x}
\langle\gamma(p,s)|J_{\mu}(x)J_{\nu}(0)^{[A]}|\gamma(p,s)\rangle \qquad
\nonumber\\
&&\sim \sum_{n} E_n^{(2)}(Q^2)
\langle\gamma(p,s)|R_n^{(2)}|\gamma(p,s)\rangle
+ \sum_{n} E_n^{(3)}(Q^2)
\langle\gamma(p,s)|R_n^{(3)}|\gamma(p,s)\rangle.\qquad
\eea
Now we calculate the virtual photon-photon forward scattering amplitude
arising from the so-called Box diagram, with $\langle e^4\rangle=
\sum_{i=1}^{N_f}e_i^4/N_f$, and $N_f$ being the number of active flavors.
\bea
&&\hspace{-0.5cm}g_1^{\gamma({\rm Box})}(x,Q^2,P^2)=
\frac{3\alpha}{\pi}N_f\langle e^4\rangle
\left[(2x-1)\ln{\frac{Q^2}{P^2}}-2(2x-1)(\ln{x}+1)\right]
\nonumber\\
&&\hspace{-0.5cm}g_2^{\gamma({\rm Box})}(x,Q^2,P^2)=
\frac{3\alpha}{\pi}N_f\langle e^4\rangle
\left[-(2x-1)\ln{\frac{Q^2}{P^2}}+2(2x-1)\ln{x}+6x-4\right].
\nonumber\\
\eea

Remarkably, the $g_2^{\gamma({\rm Box})}$ satisfies 
the Burkhardt-Cottingham sum rule \cite{BC}:
\bea
\int_0^1 dx g_2^{\gamma({\rm Box})}(x,Q^2,P^2)=0.
\eea
While it turns out that the twist-3 contribution to $g_2^\gamma$ is actually 
non-vanishing:
\bea
{\bar g}_2^\gamma&=&g_2^\gamma-g_2^{\gamma(WW)}\nonumber\\
&=&\frac{3\alpha}{\pi}N_f\langle e^4\rangle
\,\left[(2x-2-\ln{x})\ln{\frac{Q^2}{P^2}}
-2(2x-1)\ln{x}+2(x-1)+{\ln}^2{x}\right].\nonumber\\
\eea

\begin{figure}[t]
\begin{center}
\vspace{-2cm}
\includegraphics[height=12cm]{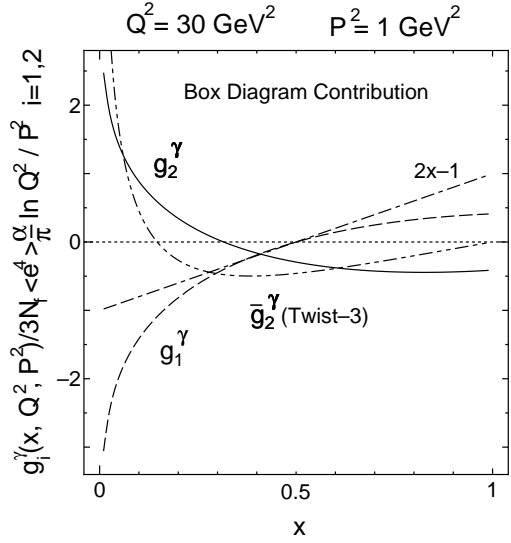}
\end{center}
\vspace{-3cm}
\caption{The Box-diagram contributions to $g_1^\gamma(x,Q^2,P^2)$ 
(dashed), $g_2^\gamma(x,Q^2,P^2)$ (solid) and 
${\bar g}_2^\gamma(x,Q^2,P^2)$ (dash-2dotted) for 
$Q^2=30$ GeV$^2$ and $P^2=1$ GeV$^2$ for $N_f=3$.}
\end{figure}
We have plotted the $g_1^\gamma$, $g_2^\gamma$ and ${\bar g}_2
^\gamma$ as a function of $x$ for the virtual photon target,
where $Q^2=30$ GeV$^2$ and $P^2=1$ GeV$^2$ for $N_f=3$ in Fig.1.

Now the OPE reads in more details 
\bea
i\int d^4x e^{iq\cdot x}T(J_\mu(x)J_\nu(0))^A
&=&-i\epsilon_{\mu\nu\lambda\sigma}q^\lambda
\sum_{n=1,3,\cdots}\left(\frac{2}{Q^2}\right)^n
q_{\mu_1}\cdots q_{\mu_{n-1}}\nonumber\\
&&\times
\left\{
\sum_i E_{i(2)}^n R_{i(2)}^{\sigma\mu_1\cdots\mu_{n-1}}
+\sum_i E_{i(3)}^n R_{i(3)}^{\sigma\mu_1\cdots\mu_{n-1}}
\right\}\nonumber\\
\eea
For the twist-2 quark operator we have 
\bea
R_{q(2)}^{\sigma\mu_1\cdots\mu_{n-1}}=
i^{n-1}\overline{\psi}\gamma_5\gamma^{\{\sigma}D^{\mu_1}
\cdots D^{\mu_{n-1}\}}\psi -\mbox{traces}
\eea
where $\{\quad\}$ is the total symmetrization with respect to 
the Lorentz indices, and the twist-2 photon operator turns out to be
\bea
R_{\gamma(2)}^{\sigma\mu_1\cdots\mu_{n-1}}=
\frac{1}{4}i^{n-1}
{\epsilon^{\{\sigma}}_{\alpha\beta\gamma}F^{\alpha\mu_1}
\partial^{\mu_2}\cdots\partial^{\mu_{n-1}\}}F^{\beta\gamma}
-\mbox{traces}
\eea
The matrix elements for these operators read
\bea
&&\langle\gamma(p,s)|R_{i(2)}^{\sigma\mu_1\cdots\mu_{n-1}}|
\gamma(p,s)\rangle=a_{i(2)}^n
s^{\{\sigma}p^{\mu_1}\cdots p^{\mu_{n-1}\}}\nonumber\\
&&
s^{\{\sigma}p^{\mu_1}\cdots p^{\mu_{n-1}\}}=
\frac{1}{n}\left[
s^{\sigma}p^{\mu_1}\cdots p^{\mu_{n-1}}\right.\nonumber\\
&&\left.\hspace{3cm}+s^{\mu_1}p^{\sigma}p^{\mu_2}\cdots p^{\mu_{n-1}}
\cdots +s^{\mu_{n-1}}p^{\mu_1}
\cdots p^{\mu_{n-2}}p^\sigma\right]
\eea
For the photon operator, $a_{\gamma(2)}^n=1$. 
While for the twist-3 quark operators we find
\bea
R_{q(3)}^{\sigma\mu_1\cdots\mu_{n-1}}=
i^{n-1}\overline{\psi}\gamma_5\gamma^{[\sigma,}D^{\{\mu_1]}
\cdots D^{\mu_{n-1}\}}\psi -\mbox{traces}
\eea
where $[\ ,\ ]$ denotes anti-symmetrization with respect to the
indices and the twist-3 photon operators are given by
\bea
R_{\gamma(3)}^{\sigma\mu_1\cdots\mu_{n-1}}=
\frac{1}{4}i^{n-1}
{\epsilon^{[\sigma,}}_{\alpha\beta\gamma}F^{\alpha\{\mu_1]}
\partial^{\mu_2}\cdots\partial^{\mu_{n-1}\}}F^{\beta\gamma}
-\mbox{traces}
\eea
The resulting matrix elements are
\bea
&&\langle\gamma(p,s)|R_{i(3)}^{\sigma\mu_1\cdots\mu_{n-1}}|
\gamma(p,s)\rangle=a_{i(3)}^n
s^{[\sigma,\{}p^{\mu_1]}p^{\mu_2}\cdots p^{\mu_{n-1}\}}\nonumber\\
&&
s^{[\sigma,\{}p^{\mu_1]}p^{\mu_2}\cdots p^{\mu_{n-1}\}}
=\left[\frac{n-1}{n}s^{\sigma}p^{\mu_1}\cdots p^{\mu_{n-1}}
-\frac{1}{n}\sum_{j=1}^{n-1}s^{\mu_j}p^{\mu_1}\cdots p^\sigma
\cdots p^\mu_{n-1}\right]\nonumber\\
\eea
For photon operator, we also have $a_{\gamma(3)}^n=1$.

The $n$-th moments of the photon structure functions $g_1^\gamma$
and $g_2^\gamma$ are given by
\bea
&&\int_0^1 dx x^{n-1}g_1^{\gamma}(x,Q^2,P^2)=
\sum_i a_{i(2)}^n E_{i(2)}^n(Q^2)\\
&&\int_0^1 dx x^{n-1}g_2^{\gamma}(x,Q^2,P^2)=\frac{n-1}{n}
\left[-\sum_i a_{i(2)}^n E_{i(2)}^n(Q^2)
+\sum_i a_{i(3)}^n E_{i(3)}^n(Q^2)\right]\nonumber\\
\eea
Therefore in the general framework of the OPE we conclude that
the Burkhardt-Cottingham sum rule holds \cite{BC}:
\bea
\int_0^1 dx g_2^{\gamma}(x,Q^2,P^2)=0
\eea
The mixing of quark gluon twist-3 operators are very complicated.
Thus $g_2^\gamma$ with QCD effects can be treated with operator 
mixing problem of twist-3 operators.

Pure QED effects can be studied through operator mixing between quark
and photon operators. This mixing arises from the triangular diagrams.
\bea
&&\hspace{-0.5cm}\langle\gamma(p,s)|R_{q(2)}^{\sigma\mu_1\cdots\mu_{n-1}}|
\gamma(p,s)\rangle=\frac{\alpha}{4\pi}
\left(-\frac{1}{2}K_n^{q(2)}\ln{\frac{P^2}{\mu^2}}
+A_n^{q(2)}\right)
s^{\{\sigma}p^{\mu_1}\cdots p^{\mu_{n-1}\}}\nonumber\\
&&\hspace{-0.5cm}\langle\gamma(p,s)|R_{q(3)}^{\sigma\mu_1\cdots\mu_{n-1}}|
\gamma(p,s)\rangle=\frac{\alpha}{4\pi}
\left(-\frac{1}{2}K_n^{q(3)}\ln{\frac{P^2}{\mu^2}}
+A_n^{q(3)}\right)
s^{[\sigma,\{}p^{\mu_1]}p^{\mu_2}\cdots p^{\mu_{n-1}\}}\nonumber\\
\eea
where the mixing anomalous dimensions are given by
\bea
K_n^{q(2)}=24N_f\langle e^4\rangle\frac{n-1}{n(n+1)},\quad
K_n^{q(3)}=-24N_f\langle e^4\rangle\frac{1}{n(n+1)}
\eea
The coefficient functions are given by
\bea
E_{q(2,3)}^n(1,{\bar g}(Q))=1+\cdots, \quad
E_{\gamma(2,3)}(1,{\bar g}(Q))=\frac{\alpha}{4\pi}\delta_\gamma
B_{\gamma(2,3)}^n
\eea
Thus we obtain
\bea
\int_0^1 dx x^{n-1}g_1^{\gamma(Box)}(x,Q^2,P^2)&=&\frac{3\alpha}{\pi}N_f
\langle e^4\rangle\frac{n-1}{n(n+1)}\ln\frac{Q^2}{P^2}
+\frac{\alpha}{4\pi}(
A_n^{q(2)}+\delta_\gamma B_{\gamma(2)}^n)\nonumber\\
\int_0^1 dx x^{n-1}g_2^{\gamma(Box)}(x,Q^2,P^2)
&=&-\frac{3\alpha}{\pi}N_f
\langle e^4\rangle\frac{n-1}{n(n+1)}\ln\frac{Q^2}{P^2}\nonumber\\
&&\hspace{-0.8cm}+\frac{n-1}{n}\frac{\alpha}{4\pi}(
-A_n^{q(2)}-\delta_\gamma B_{\gamma(2)}^n
+A_n^{q(3)}+\delta_\gamma B_{\gamma(3)}^n)
\eea
This is nothing but the Box-diagram contribution.
Here we note that there exists the twist-3 effects even at the
pure QED level.
Here we also note that $g_1^\gamma+g_2^\gamma$ does
not have $\ln{Q^2/P^2}$ behavior. 

\section{Concluding Remarks}

We have investigated the virtual photon's spin structure functions
$g_1^\gamma(x,Q^2,P^2)$ and $g_2^\gamma(x,Q^2,P^2)$ in the 
kinematical region $\Lambda^2 \ll P^2 \ll Q^2$. The 1st moment of 
$g_1^\gamma$ is non-vanishing which leads to a sum rule. The polarized
quark and gluon distributions are computed to the NLO by the 
perturbative method, but these distributions are factorization-scheme
dependent.
The twist-3 effects do exist in $g_2^\gamma$ for the virtual photon 
target. Here we have discussed the OPE analysis for the pure QED
effects corresponding to the Box-diagrams.
The full QCD analysis is now under investigation. Finally we can
also discuss the positivity constraints on the polarized and 
unpolarized photon structure functions \cite{SSU}.

\bigskip
{\small One of the authors (T.U.) would like to express his
sincere thanks to the Organizers of the ASI, especially to 
Prof. Finger for the wonderful organization and the 
hospitality during the School.}

\end{document}